\documentclass[prd,twocolumn,unsortedaddress,superscriptaddress]{revtex4-1}
\usepackage{graphicx}
\usepackage{epsfig,color}
\usepackage{amsmath,amssymb,eufrak}
\usepackage{amsthm}
\usepackage{enumerate}
\usepackage{hyperref}
\usepackage{amsxtra,amsfonts,bm}

\begin{document}

\title{Local Manipulation and Measurement of Nonlocal Many-Body Operators in Lattice Gauge Theory Quantum Simulators}

\date{\today}

\author{Erez Zohar}
\address{Racah Institute of Physics, The Hebrew University of Jerusalem, Jerusalem 91904, Givat Ram, Israel.}

\begin{abstract}
Lattice Gauge Theories form a very successful framework for studying nonperturbative gauge field physics, in particular in Quantum Chromodynamics. Recently, their quantum simulation on atomic and solid-state platforms has been discussed, aiming at overcoming some of the difficulties still faced by the conventional approaches (such as the sign problem and real time evolution). While the actual implementations of a lattice gauge theory on a quantum simulator may differ in terms of the simulating system and its properties, they are all directed at studying similar physical phenomena, requiring the measurement of nonlocal observables, due to the local symmetry of gauge theories. In this work, general schemes for measuring such nonlocal observables (Wilson loops and mesonic string operators) in general lattice gauge theory quantum simulators that are based merely on local operations are proposed.
\end{abstract}

\maketitle

\section{Introduction}
Lattice gauge theories (LGTs) reformulate initially continuous gauge fields (which carry the forces within the standard model of particle physics)
on discrete spacetimes \cite{wilson_confinement_1974,kogut_introduction_1979} or spaces \cite{kogut_hamiltonian_1975}. This allows one to perform very fruitful numerical calculations, suitable for the nonperturbative nature of the relevant theories, such as and first of all QCD (Quantum Chromodynamics), the theory of the strong force). While being extremely successful for a variety of physical properties (e.g. the hadronic spectrum \cite{flag_working_group_review_2014}), they still face some difficulties due to the computation method: Monte-Carlo path integration in Wick-rotated, Euclidean spacetimes. One issue is the well-known sign problem \cite{troyer_computational_2005}, arising in scenarios with finite fermionic densities (such as in several interesting regimes of the QCD phase diagram \cite{mclerran_physics_1986,fukushima_phase_2011}); the other one is the inability to directly describe unitary real-time evolution.

One possible way to overcome these problems is quantum simulation \cite{feynman_simulating_1982}, where the system of interest is mapped to another quantum system that is controllable in the lab, serving as a table-top simulation of otherwise inacceseible physics. Quantum simulation of lattice gauge theories \cite{wiese_ultracold_2013,zohar_quantum_2016,dalmonte_lattice_2016} has been proposed by several research groups in the last years \cite{zohar_confinement_2011,banerjee_atomic_2012,tagliacozzo_optical_2013,zohar_simulating_2012,zohar_simulating_2013,tagliacozzo_simulation_2012,zohar_topological_2013,zohar_quantum_2013,banerjee_atomic_2013,hauke_quantum_2013,marcos_superconducting_2013,stannigel_constrained_2014,marcos_two-dimensional_2014,bazavov_gauge_2015,kuno_formulation_2014,notarnicola_discrete_2015,mezzacapo_non-abelian_2015,zohar_digital_2017,zohar_digital_2017-1,Kasper_Implementing_2017,dutta_toolbox_2017,gonzalez-cuadra_quantum_2017,bender_digital_2018,zache_quantun_2018,rico_so_2018,surace_lattice_2019,celi_emerging_2019,davoudi_towards_2019}, addressing various gauge groups and models. The proposed simulating system has so far mostly been cold atoms in optical lattices \cite{jaksch_cold_1998,bloch_many-body_2008,lewenstein_ultracold_2012}, but other simulating systems have been proposed for the purpose as well. Several pioneering experiments have  already been carried out as well \cite{martinez_real-time_2016,kokail_self_2019,mil_realizing_2019,aidelsburger_floquet_2019}, demonstrating the great potential of the quantum simulation approach. Recently, quantum Computer algorithms for lattice gauge theories have also been studied  \cite{klco_quantum_2018,kaplan_gauss_2018,preskill_simulating_2018,stryker_oracles_2019,klco_su_2019}.

Regardless of how the simulation is performed, the local symmetry which is in the core of lattice gauge theories poses a strong restriction on the relevant physical observables: they must be gauge invariant, otherwise they vanish.  For that reason, they introduce some physically relevant nonlocal many-body operators, such as the Wilson loop \cite{wilson_confinement_1974} or mesonic string operators. Manipulating and measuring such operators may be problematic, unless some special procedure is used. In this work, we suggest ways to do exactly that: map the information stored in such nonlocal many-body observables to local ancillary degrees of freedom, using sequences of local and simple two-body interactions of the ancilla with the physical degrees of freedom. The ancilla can be manipulated and measured locally. The general method presented here can be used in various quantum simulations and quantum computer algorithms of lattice gauge theories, and allow one to efficiently extract highly relevant information from the quantum state of the simulator.

The paper begins with a review of some essential lattice gauge theory background. Then we turn to a general formulation of the scheme, for arbitrary gauge groups, and finally - demonstrate it for the simple case of quantum simulators of $\mathbb{Z}_2$ lattice gauge theories with matter \cite{gazit_emergent_2017}, which, thanks to their simplicity in terms of Hilbert spaces and degrees of freedom, have been studied recently \cite{zohar_digital_2017,aidelsburger_floquet_2019}, as prototypes of quantum simulators of more complicated models. Throughout the paper, summation of repeated indices is assumed.

\section{Lattice Gauge Theory Background}

\subsection{The Hilbert Space}
A general lattice gauge theory contains two types of degrees of freedom: matter fields, which reside on the vertices of the lattice (labelled by $\mathbf{x} \in \mathbb{Z}^d$ for a $d$ dimensional spatial lattice), and gauge fields, which reside on its links (labelled by $\left(\mathbf{x};i\right)$ - a pair of a vertex and a direction $i \in \left\{1,...,d\right\}$). We denote a unit vector in direction $i$ with $\hat{\mathbf{e}}_i$.

The relevant Hilbert space is a subspace of the tensor product of a fermionic Fock space on all the vertices, describing the matter, and the gauge field Hilbert space which is, itself, a tensor product of local spaces on the links. Let $G=\left\{g\right\}$ be our gauge group. Then, the local Hilbert spaces on the links may be spanned by the \emph{group elements basis}, whose elements $\left\{\left|g\right\rangle\right\}_{g\in G}$ are labelled by the different gauge group elements. In the finite case (e.g., $G=\mathbb{Z}_N$), the local Hilbert space dimension is the number of group elements, and it is an orthonormal basis: $\left\langle g' | g \right\rangle = \delta_{g,g'}$; if $G$ is infinite (e.g. a compact Lie group such as $U(N)$ or $SU(N)$), so is the link's Hilbert space dimension, and an orthogonality relation of the form $\left\langle g' | g \right\rangle = \delta\left(g,g'\right)$, with a distribution defined by the group's measure, holds instead. We define unitary transformation operators, parameterized by the gauge group elements, $\Theta^R_g$ and $\Theta^L_g$, responsible for right and left group operations, respectively:
\begin{equation}
\Theta^R_g \left|h\right\rangle = \left|hg^{-1}\right\rangle \quad ; \quad 
\Theta^L_g \left|h\right\rangle =  \left|g^{-1}h\right\rangle
\end{equation}

The matter fermions on a vertex $\mathbf{x}$ are created by the spinor components $\psi^{\dagger}_m\left(\mathbf{x}\right)$, belonging to a multiplet of a fixed representation (e.g. the fundamental one). We define a unitary transformation operator, parametrized by the gauge group elements, $\theta_g$, for the matter:
\begin{equation}
\theta_g \psi^{\dagger}_m \theta^{\dagger}_g = \psi^{\dagger}_n D^{j}_{nm}\left(g\right) 
\end{equation}
where $D^j\left(g\right)$ is the $j$ irreducible representation of $g$. Here, for simplicity and following the usual choice for quantum simulation , we use a staggered fermionic picture \cite{susskind_lattice_1977}, in which the two different sublattices correspond to particles and antiparticles. For that, we define a generalized transformation operator \cite{zohar_formulation_2015},
\begin{equation}
\check{\theta}_g\left(\mathbf{x}\right) = \begin{cases}
   \theta_g\left(\mathbf{x}\right) & \mathbf{x} \text{ is on the even sublattice} \\
   \theta_g\left(\mathbf{x}\right) \det\left(g^{-1}\right)       & \mathbf{x} \text{ is on the odd sublattice.}
\end{cases}
\end{equation}
where $\det\left(g^{-1}\right)$ is the determinant of the matter's irreducible representation of $g$. When the group is Abelian (e.g. $\mathbb{Z}_N$ or $U(1)$), $\Theta^R_g =\Theta^L_g \equiv \Theta_g$.

Gauge invariance is invariance under all local transformations of the form
\begin{equation}
\hat{\Theta}_g\left(\mathbf{x}\right)=\underset{i=1,...,d}{\prod}\left(\Theta^L_g\left(\mathbf{x};i\right) \Theta^{R\dagger}_g\left(\mathbf{x}-\hat{\mathbf{e}}_i;i\right)\right)\check{\theta}^{\dagger}_g\left(\mathbf{x}\right)
\end{equation}
- that is, a gauge invariant state $\left|\psi\right\rangle$ satisfies
\begin{equation}
\hat{\Theta}_g\left(\mathbf{x}\right)\left|\psi\right\rangle = \left|\psi\right\rangle, \quad \forall \mathbf{x} \in \mathbb{Z}^d
\end{equation}
(disregarding static charges, which are not discussed here), and a gauge invariant Hamiltonian $H$ satisfies
\begin{equation}
\left[\hat{\Theta}_g\left(\mathbf{x}\right),H\right]=0, \quad \forall \mathbf{x} \in \mathbb{Z}^d
\end{equation}
implying that, indeed, the \emph{physical} or gauge invariant states $\left|\psi\right\rangle$ form only a subspace of the product space of matter and gauge fields. The dynamics is conventionally given by the Kogut-Susskind Hamiltonian \cite{kogut_hamiltonian_1975,kogut_introduction_1979,zohar_formulation_2015}.

\subsection{Gauge Invariant Operators}
When studying a gauge theory, only operators that are gauge invariant are relevant physical observables. These can be local operators such as the total number of fermions on a vertex,
\begin{equation}
n\left(\mathbf{x}\right) = \psi^{\dagger}_m\left(\mathbf{x}\right)\psi_m\left(\mathbf{x}\right)
\end{equation}
or electric field strength operators, local on the links, that are diagonal in the so-called \emph{representation} basis, dual to the group element one \cite{zohar_formulation_2015} that we do not discuss here. As both types of operators are local, it is more than reasonable to assume that their measurement in a quantum simulator is not a complicated task, and only depends on properties of the implementation and experimental settings. It is the other type of operators, the nonlocal ones, whose manipulation and measurement may pose a challenge, and they are the ones addressed below.

Before introducing such operators, we have to define the \emph{group element operators}, which act on the gauge field link Hilbert spaces. They serve as the connections that maintain local gauge invariance, and thus strings thereof are present in nonlocal gauge invariant operators, giving them a many-body nature. On a link Hilbert space we define $U^{j}_{mn}$, a square unitary matrix of operators, whose size is the dimension of $j$. Its elements are operators acting on the link Hilbert space:
\begin{equation}
U^{j}_{mn} = \int dg \left|g\right\rangle \left\langle g \right| D^{j}_{mn}\left(g\right)
\label{Udef}
\end{equation}
where the integration is replaced by a sum for a finite group. One can see that all the elements of $U^j$ are simultaneously diagonalizable, and thus commute. Therefore, matrix operations on $U^j_{mn}$ are well defined, just as if it were a matrix of numbers \cite{zohar_formulation_2015,zohar_digital_2017-1}.
 Their eigenstates of the matrix elements $U^j_{mn}$ are  group element states: $U^{j}_{mn}  \left|g\right\rangle =  D^{j}_{mn}\left(g\right)\left|g\right\rangle$. In the following, the index $j$ will be omitted and $U_{mn}$ will be used, for the representation used for the matter (mostly the fundamental). 

\begin{figure}
	\includegraphics[width=0.7\columnwidth]{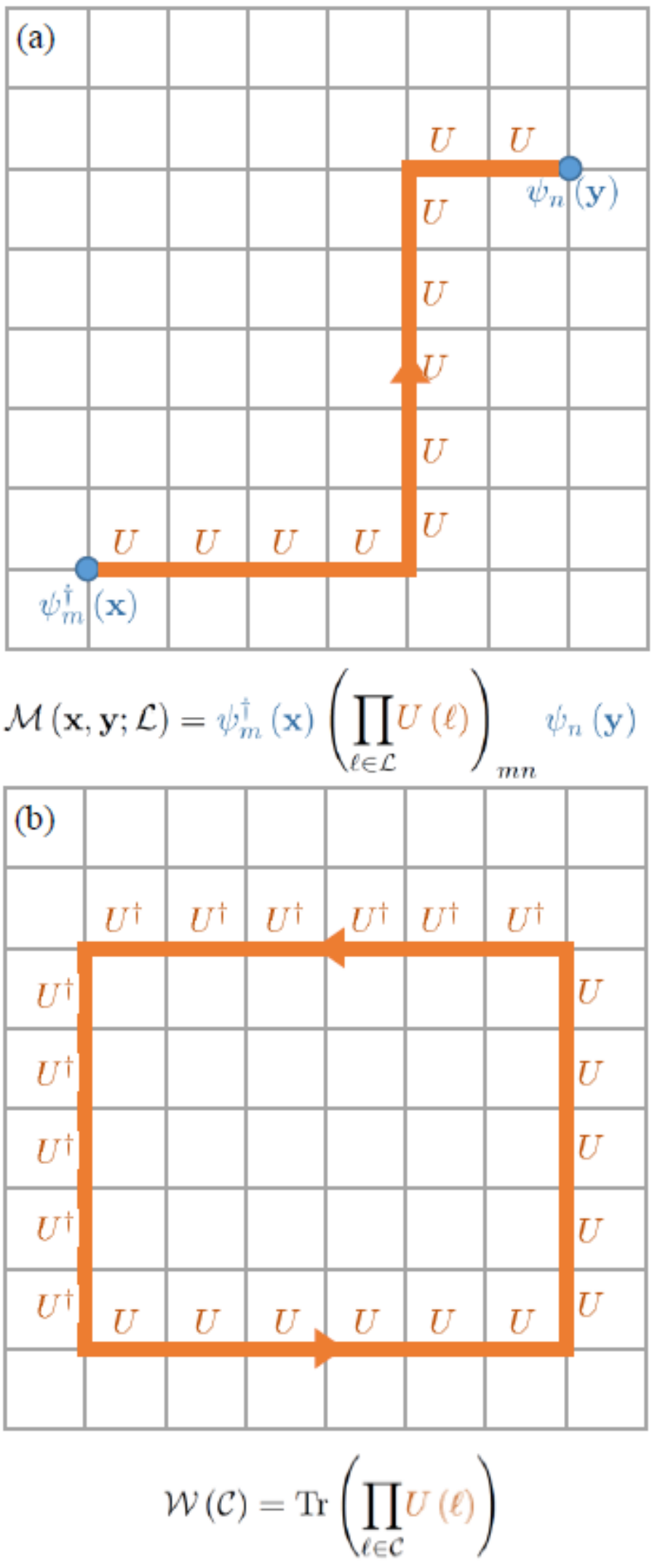}
	\caption{(a) A mesonic string operator, as in (\ref{Mdef}). (b) A Wilson loop, as in (\ref{Wdef}).}
	\label{Fig1}
\end{figure}

In a gauge theory, a fermionic two-point correlator of the form $\left\langle \psi^{\dagger}_m\left(\mathbf{x}\right) \psi_m\left(\mathbf{y}\right)\right\rangle$ will vanish if $\mathbf{x}\neq\mathbf{y}$, since the operator $\psi^{\dagger}_m\left(\mathbf{x}\right) \psi_m\left(\mathbf{y}\right)$ is not gauge invariant. This is fixed by connecting the operators by a string of group element operators, along some path that connects the two vertices, defining the \emph{mesonic string operator}:
\begin{equation}
\mathcal{M}\left(\mathbf{x},\mathbf{y};\mathcal{L}\right) = \psi^{\dagger}_m\left(\mathbf{x}\right) \left(\underset{\ell \in \mathcal{L}}{\prod}U\left(\ell\right)\right)_{mn}\psi_n\left(\mathbf{y}\right)
\label{Mdef}
\end{equation}
where $\mathbf{x}\neq\mathbf{y}$, $\mathcal{L}$ is some path from $\mathbf{x}$ to $\mathbf{y}$, and $\underset{\ell \in \mathcal{L}}{\prod}U\left(\ell\right)$ is an ordered product of the matrices $U$ along $\mathcal{L}$ (in which some $U$ operators have to be replaced by $U^{\dagger}$ according to the gauge invariant orientation; See Fig. \ref{Fig1}). This is the gauge invariant operator whose expectation value replaces fermionic two-point functions such as $\left\langle \psi^{\dagger}_m\left(\mathbf{x}\right) \psi_m\left(\mathbf{y}\right)\right\rangle$ (and their choice is not unique because there are many possible paths connecting the endpoints). The main point is that introduction of gauge invariance increases the complexity, and turns the correlators into many-body objects, that do not only depened on the endpoints, but also on gauge field degrees of freedom along the path. Almost all the mesonic operators commute: only ones that share an endpoint (either one or both) do not. Therefore, almost any two meson operators may be simultaneously measured in theory.

Another important operator is the Wilson loop, which is defined also for a pure gauge theory: the trace of the ordered product of group elements operators along some closed curve $C$, mostly a rectangle \cite{wilson_confinement_1974}. The decay law of a rectangular planar Wilson loop in the thermodynamic limit of a pure gauge theory determines whether static charges are confined (area law decay) or deconfined (perimeter law decay) \cite{wilson_confinement_1974,polyakov_quark_1977,fradkin_order_1978,polyakov_gauge_1987}. It is defined by
\begin{equation}
\mathcal{W}\left(\mathcal{C}\right) = \text{Tr}\left(\underset{\ell \in \mathcal{C}}{\prod}U\left(\ell\right)\right) \equiv \text{Tr}\left(W\left(\mathcal{C}\right)\right)
\label{Wdef}
\end{equation}
where the trace guarantees gauge invariance, for an ordered matrix product along $\mathcal{C}$, as long as the right orientation ($U$ or $U^{\dagger}$) is chosen (see Fig. \ref{Fig1}). All different Wilson loops commute, and therefore may be, theoretically, measured simultaneously.

The shortest mesonic operators appear in the conventional Hamiltonian part that couples the gauge field and the matter,
\begin{equation}
H_{\text{GM}} = \underset{\mathbf{x},i}{\sum}\left(\lambda_{\text{GM}}\left(\mathbf{x},i\right)
\psi^{\dagger}_m\left(\mathbf{x}\right)U_{mn}\left(\mathbf{x};i\right)\psi_n\left(\mathbf{x}+\hat{\mathbf{e}}_i\right) + H.c.\right)
\end{equation}
The shortest Wilson loop - around one unit square (plaquette) of the lattice, forms the magnetic four-body interaction (plaquette interaction) out of which the magnetic part of the Kogut-Susskind pure gauge Hamiltonian is constructed \cite{kogut_hamiltonian_1975}:
\begin{widetext}
\begin{equation}
H_{\text{B}}=-\lambda_{\text{B}}\underset{\mathbf{x},i,j}{\sum}\left(\text{Tr}\left(
U\left(\mathbf{x},i\right)U\left(\mathbf{x}+\hat{\mathbf{e}}_i,j\right)U^{\dagger}\left(\mathbf{x}+\hat{\mathbf{e}}_j,i\right)
U^{\dagger}\left(\mathbf{x},j\right)\right) + H.c.\right)
\end{equation}
\end{widetext}
And thus, being able to measure such operators is also crucial for the measurement of energy, or Hamiltonian expectation value.

Since the mesonic strings and the Wilson loops are gauge invariant, they are also useful in the creation and manipulation of gauge invariant states. It is common to represent any gauge invariant state as the outcome of acting with such operators on the so-called \emph{strong-limit eigenstates} - states with no fermionic excitations in which the gauge field is in a product state of singlet states $\left|0\right\rangle$ (also called zero electric field states, for which $\Theta^R_g\left|0\right\rangle=\Theta^L_g\left|0\right\rangle=\left|0\right\rangle$) \cite{kogut_hamiltonian_1975,kogut_introduction_1979}.

\section{Manipulating and Measuring the Non-local Observables in a Quantum Simulator}

After having reviewed the necessary background ingredients, we can now move on to the schemes for measuring expectation values of Wilson loops and mesonic strings, as well as their manipulation for state preparation. We do not assume anything on the nature of the quantum simulator and the simulating scheme or system, besides the following two assumptions:
\begin{enumerate}
	\item We work with a given state, $\left|\psi\right\rangle$, which is gauge invariant. It is either the result of some time evolution of the simulator, at an instance of time in which we wish to perform a measurement, or awaiting some time evolution before which, in the current time, we wish to change it in a gauge invariant way. This way or the other, we assume that we obtain it when the gauge field dynamics is completely switched off, and it is ready for manipulation.
	\item The physical Hilbert spaces are either the exact ones needed for the simulation (feasible simulator for finite groups, idealistic scenario for infinite ones) or a truncation in which the gauge group is restricted to a subgroup (as in the stator scheme of \cite{zohar_digital_2017-1}). This is important because, as will shortly become clear, the scheme requires to use either the original group elements $U_{mn}$, or another unitary approximation of those. For example, this is satisfied by an approximation of $U(1)$ with $\mathbb{Z}_N$ \cite{horn_hamiltonian_1979}, which is a subgroup and keeps the group structure, and not electric field truncation as in \cite{zohar_simulating_2012,zohar_quantum_2013}.
\end{enumerate}

The scheme is inspired by the stator formalism \cite{reznik_remote_2002,zohar_half_2017} and its application to quantum simulation of lattice gauge theories \cite{zohar_digital_2017,zohar_digital_2017-1,bender_digital_2018}. Thus, using a similar concept for the quantum simulation will guarantee that the simulating system is properly equipped and capable of carrying out the unitary operations required for the schemes to be described. Nevertheless, it is possible to realize them in other types of quantum simulators as well, depending on the platform and the experimental setting.

\subsection{Wilson Loop Actions and Measurements}

We begin our discussion with Wilson loops. The key point is that the operator we discuss, $\mathcal{W}$, is obtained from tracing a product of group element operators, $W$	 (\ref{Wdef}). Since a product of group elements is, itself, a group element, a product of group element operators, such as $W$, is a group element operator. Therefore, in order to store information of the Wilson loop, we do not need the product space of all the link Hilbert spaces along its path: one such Hilbert space is enough for storing this information, no matter how long the loop is. When reducing to the smallest case, of the plaquette, this is simply the known fact that the magnetic field through a plaquette resides in a Hilbert space that is identical to those of the four vector potential spaces on the plaquette's links.

For storing the Wilson loop's information locally, we need  ancillary degree of freedom, whose Hilbert space is mathematically identical to those on the links. For example, when discussing a $\mathbb{Z}_2$ lattice gauge theory, where the links are occupied by two level systems, or qubits, we will need an auxiliary qubit.
The ancilla should be movable in a controlled way. For extracting the Wilson loop's data, the ancilla has to be taken a long the closed path, and interact with each of the links alone, in a sequential, ordered way. In these interactions it will collect the information on the loop's state. At the end, the Wilson loop's information will be stored at the ancilla, which can then be taken away and measured locally.

\subsubsection{Key Idea and Ingredients}

The process begins with a product state of the physical system and the ancilla:
\begin{equation}
\left|\Psi\right\rangle = \left|\psi\right\rangle \otimes \left|\tilde{e}\right\rangle
\end{equation}
the physical system is in the state of interest $\left|\psi\right\rangle$, and the ancilla (whose states and operators  are denoted with a $\sim$ here and below) is prepared initially in the group element state corresponding to the identity.

We define the unitary operation $\mathcal{U}_\text{W}\left(\ell\right)$, between the ancilla and a link $\ell$,
\begin{equation}
\mathcal{U}_{\text{W}}\left(\ell\right) = \int dg \left|g\right\rangle\left\langle g\right|_{\ell} \otimes \tilde{\Theta}^{L\dagger}_g
\label{UWldef}
\end{equation}
Out of those operators, we define the Wilson loop entangling operator, for the loop $\mathcal{C}$,
\begin{equation}
\mathcal{U}_{\text{W}}\left(\mathcal{C}\right) = \mathcal{P}\left( \underset{\ell \in \mathcal{C}}{\prod} \mathcal{U}_{\text{W}}\left(\ell\right)\right)
\label{UWCdef}
\end{equation}
where $\mathcal{P}$ stands for path ordering: the local unitaries $\mathcal{U}_{\text{W}}\left(\ell\right)$ are multiplied in an order that matches that of the Wilson loop definition for $\mathcal{C}$ (\ref{Wdef}), which includes replacing by $\mathcal{U}^{\dagger}_{\text{W}}\left(\ell\right)$ when required by the orientation. The starting point is not important since eventually we will only care about a trace along $\mathcal{C}$. We define a map $S_{\text{W}}$ from the physical Hilbert space to the product of physical and ancilla spaces (such a map is called a \emph{stator} \cite{reznik_remote_2002,zohar_half_2017}),
\begin{equation}
S_{\text{W}}\left(\mathcal{C}\right) = \mathcal{U}_{\text{W}}\left(\mathcal{C}\right) \left|\tilde{e}\right\rangle
\end{equation}
which takes the form
\begin{equation}
S_{\text{W}}\left(\mathcal{C}\right) = \int \underset{\ell\in\mathcal{C}}{\prod} dg\left(\ell\right) \left|\left\{g\right\}\right\rangle \left\langle \left\{g\right\} \right| 
\otimes\left| \tilde{\mathcal{G}}\right\rangle
\label{SW}
\end{equation}
where $\underset{\ell\in\mathcal{C}}{\prod} dg\left(\ell\right)$ integrates (or sums) over all the possible gauge field configurations along $\mathcal{C}$,
$\left|\left\{g\right\}\right\rangle$ is a product state of all the links in $\mathcal{C}$ corresponding to a configuration, and
 $\mathcal{G}=g_1 g_2 \cdots$ is the oriented product of all the group elements along $\mathcal{C}$. It is straightforward to see that the stator $S_{\text{W}}$ satisfies the eigenoperator relation \cite{zohar_half_2017} for group element operators,
 \begin{equation}
 \tilde{U}_{mn} S_{\text{W}}\left(\mathcal{C}\right) = S_{\text{W}}\left(\mathcal{C}\right) W_{mn}\left(\mathcal{C}\right)
 \end{equation}
 That is, a local group element operator action on the ancilla intertwines through $S_{\text{W}}\left(\mathcal{C}\right)$ as the matrix product $W\left(\mathcal{C}\right)$ along the path $\mathcal{C}$. In particular, this applices to the trace:
 \begin{equation}
\text{Tr}\left(\tilde{U}\right) S_{\text{W}}\left(\mathcal{C}\right) = S_{\text{W}}\left(\mathcal{C}\right) \mathcal{W}\left(\mathcal{C}\right)
\end{equation} 
  This implies that measuring the ancilla after the stator is generated is equivalent to measuring the Wilson loop:
  \begin{equation}
  \left\langle \psi \right| \mathcal{W}\left(\mathcal{C}\right) \left| \psi \right\rangle=
  \left\langle \Psi \right| 
  \mathcal{U}^{\dagger}_{\text{W}}\left(\mathcal{C}\right)
  \text{Tr}{\tilde{U}}
  \mathcal{U}_{\text{W}}\left(\mathcal{C}\right)
  \left| \Psi \right\rangle 
  \label{eow}
  \end{equation}
  with the initial product state $\left|\Psi\right\rangle =  \left| \psi \right\rangle \otimes \left| \tilde{e}\right\rangle$.
 
 \subsubsection{The Scheme}
  
  \begin{figure*}[t!]
  	\includegraphics[width=\textwidth]{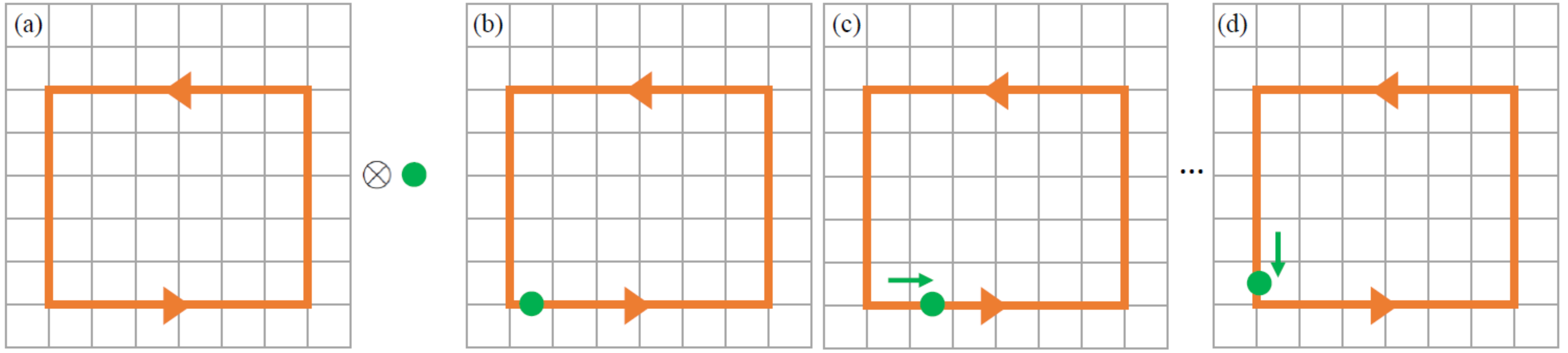}
  	\caption{The scheme for a Wilson loop: (a) the ancilla (green circle) is prepared in its initial state, in a product with the system's state. 
  		(b) the ancilla is brought to the first link of the loop, $\ell_i$ (a matter of choice), and interacts with it through $\mathcal{U}_{\text{W}}\left(\ell_i\right)$. (c) the ancilla is moved to the next link and interacts similarly with it. (d) after having interacted similarly with all the links until the last remaining one, $\mathcal{U}_{\text{W}}\left(\mathcal{C}\right)$ is achieved, and the ancilla stores the Wilson loop's information - which can be measured or used for state preparation.}
  	\label{Fig2}
  \end{figure*}
  
  The measurement/actions scheme will implement the above procedure, as follows:
  \begin{enumerate}
  	\item Prepare the ancilla in the initial state $\left|\tilde{e}\right\rangle$, giving rise to the initial product state $\left|\Psi\right\rangle =  \left| \psi \right\rangle \otimes \left| \tilde{e}\right\rangle$.
  	\item Move the ancilla slowly along the Wilson Loop path $\mathcal{C}$, in a way that realizes the sequence of unitaries $\mathcal{U}_W\left(\mathcal{C}\right)$ as defined in (\ref{UWldef}): when brought to a link $\ell$, $\mathcal{U}_W\left(\mathcal{\ell}\right)$ is realized. Such operations may be realized, for example, in cold atomic settings (optical lattices) using prescriptions given in \cite{zohar_digital_2017,zohar_digital_2017-1,bender_digital_2018} where a similar procedure is used for obtaining the gauge invariant dynamics of the quantum simulator.
  	\item When the loop is closed, the stator $S_{\text{W}}\left(\mathcal{C}\right)$ (\ref{SW}) is ready for use. The Wilson loop may be either read-out locally by measuring the ancilla, as shown in (\ref{eow}), or, alternatively, one can use it for the excitation of a loop: after entangling with the ancilla, a local action on it is equivalent to acting with the loop operator on the physical system, and the only thing left to do is to disentangle the ancilla by reversing the process:
\begin{equation}
\left(\mathcal{W}\left(\mathcal{C}\right) \left|\psi\right\rangle \right) \otimes \left|\tilde{e}\right\rangle =
\mathcal{U}_{\text{W}}^{\dagger}\left(\mathcal{C}\right)  \text{Tr}{\tilde{U}} \mathcal{U}_{\text{W}}\left(\mathcal{C}\right) \left|\Psi\right\rangle
\end{equation}  	
  \end{enumerate}
The process is shown on Fig. \ref{Fig2}.

This procedure allows one to measure the nonlocal Wilson loop, or excite a loop, using only local operations and two-body interactions. Since all Wilson loops commute, one can use this procedure to measure the expectation value of several Wilson loops, or to excite multiple loops, by using different ancillas, each corresponding to another loop.

\subsection{Mesonic String Measurements and Actions}

Next, we consider the measurement of the mesonic string operator, $\mathcal{M}\left(\mathbf{x},\mathbf{y};\mathcal{L}\right)$ (\ref{Mdef}), or a way to implement their action on a state. This will require a fermionic ancilla, movable as well, with spinor components created by $\chi^{\dagger}_m$, forming a multiplet identical to the local matter multiplet $\psi^{\dagger}_m\left(\mathbf{x}\right)$. We will first transfer the information from the matter fermions at the end of the string, $\mathbf{y}$ to the ancilla, and then move it along the path $\mathcal{L}$ and telescopically shorten the string, until it reaches the starting point $\mathbf{x}$ where acting on it or measuring it will correspond to performing the same on the whole meson, as described below.

\subsubsection{Building Blocks}

For this procedure, we have to define several unitary transformations. First, the fermionic swap operator $\mathcal{U}_S\left(\psi,\chi\right)$, that swaps the fermionic modes created by $\psi^{\dagger}$ and $\chi^{\dagger}$. We make use of the fact that fermionic number operators $n$ have only zero and one in their spectrum, just like projection operators, and thus $n$ is a projector to the occupied state, and $1-n$ is a projector to the empty one. $\mathcal{U}_S\left(\psi,\chi\right)$ should not affect a state in which none of the modes is occupied, change the excited mode if the occupation is one and change the sign (corresponding to fermionic exchange) if both modes are occupied; therefore,
\begin{equation}
\begin{aligned}
\mathcal{U}_S\left(\psi,\chi\right) & = \left(1-n_{\psi}\right)\left(1-n_{\chi}\right) \\
&+ \left( \psi^{\dagger}\chi + \chi^{\dagger}\psi\right)\left(n_{\psi}\left(1-n_{\chi}\right)+n_{\chi}\left(1-n_{\psi}\right)\right) \\
&-n_{\psi}n_{\chi}
\end{aligned}
\end{equation}
This transformation gives rise to $\mathcal{U}_S\left(\psi,\chi\right) \psi^{\dagger} \mathcal{U}^{\dagger}_S\left(\psi,\chi\right) = \chi^{\dagger}$ and 
 $\mathcal{U}_S\left(\psi,\chi\right) \chi^{\dagger} \mathcal{U}^{\dagger}_S\left(\psi,\chi\right) = \psi^{\dagger}$.
 
 The second type of a unitary is $\mathcal{U}_{\text{G}}$, which rotates a fermion with respect to a gauge field operator:
 \begin{equation}
 \mathcal{U}_{\text{G}} \left(\chi,U\right) \chi^{\dagger}_m \mathcal{U}^{\dagger}_{\text{G}} \left(\chi,U\right) =
 U^{\dagger}_{mn} \chi^{\dagger}_n
 \end{equation}
 This can be implemented via a unitary transformation, since from the definition of the group element operator (\ref{Udef}) it is clear that it is a unitary matrix of operators. In \cite{zohar_digital_2017,zohar_digital_2017-1,bender_digital_2018} explicit constructions of such transformations in several cold atomic simulators are given. In \cite{zohar_projected_2016,zohar_combining_2018,emonts_gauss_2018} a gauging transformation which couples fermions to gauge fields in a gauge invariant way is introduced; the transformation defined here, which will be used for shortening the mesonic string by removing, link-by-link, the included group element operators, is simply its inverse - a \emph{de-gauging} transformation.
 
 The last unitary we introduce is a fermionic rotation operator. Out of two fermionic modes, created by $\psi$ and $\chi$, one can construct an $SU(2)$ algebra:
 \begin{equation}
 \begin{aligned}
 S_x &= \frac{1}{2}\left(\psi^{\dagger}\chi + \chi^{\dagger}\psi\right) \\
 S_y &= \frac{i}{2}\left(\psi^{\dagger}\chi - \chi^{\dagger}\psi\right) \\
 S_z &= \frac{1}{2}\left(\psi^{\dagger}\psi - \chi^{\dagger}\chi\right) = \frac{1}{2}\left(n_{\psi} - n_{\chi}\right)
 \end{aligned}
 \end{equation}
 The transformation 
 \begin{equation}
  \mathcal{U}_{\text{R}}=\exp\left(i \pi S_y / 2\right)
  \end{equation} 
  rotates, as usual, $S_x$ to $S_z$, that is
 \begin{equation}
 \mathcal{U}_{\text{R}} \left(\psi^{\dagger}\chi + \chi^{\dagger}\psi\right) \mathcal{U}^{\dagger}_{\text{R}} = n_{\psi} - n_{\chi}
 \end{equation} 
 
 \subsubsection{The Scheme}
 
 The actual observable to be measured will be the expectation value of
 \begin{equation}
 M\left(\mathbf{x},\mathbf{y}\right) = \mathcal{M}\left(\mathbf{x},\mathbf{y}\right) + \mathcal{M}^{\dagger}\left(\mathbf{x},\mathbf{y}\right)
 \end{equation}
 A similar scheme can be applied for measuring 
  \begin{equation}
 M'\left(\mathbf{x},\mathbf{y}\right) = -i\left(\mathcal{M}\left(\mathbf{x},\mathbf{y}\right) - \mathcal{M}^{\dagger}\left(\mathbf{x},\mathbf{y}\right)\right)
 \end{equation}
 and then
 \begin{equation}
 \left\langle \mathcal{M}\right\rangle = \frac{1}{2}\left\langle M + M'\right\rangle
 \end{equation}
 
 We let $L$ be the length of $\mathcal{L}$, label the oriented links along it, from $\mathbf{x}$ to $\mathbf{y}$ with $\ell=1,...,L$, and rewrite $\mathcal{M}$ as
 \begin{equation}
\mathcal{M}\left(\mathbf{x},\mathbf{y};\mathcal{L}\right) = \psi^{\dagger}_m\left(\mathbf{x}\right) \left(\underset{\ell=1,...,L }{\prod}U\left(\ell\right)\right)_{mn}\psi_n\left(\mathbf{y}\right)
 \end{equation}
 
 The scheme for the mesons is the following:
 \begin{enumerate}
 	\item Introduce the ancillary fermionic modes $\chi$ in an empty state; formally, we embed the physical Hilbert space in a larger one that includes the ancilla, and the physical state $\left|\psi\right\rangle$ is lifted to 
 	\begin{equation}
	\left|\Psi\right\rangle = \left|\psi\right\rangle \otimes \left|\Omega_{\chi}\right\rangle 
 	\end{equation}
 	where $\left|\Omega_{\chi}\right\rangle$ is the Fock vacuum of the $\chi$ modes \footnote{In general, some ordering has to be defined in order to give meaning to the above fermionic tensor product; however, this is not a problem in the current case where the ancilla state that we multiply with is the vacuum.}.  
 	\item Bring the ancilla close to $\mathbf{y}$. Bringing an ancilla which contains no fermions should be understood as placing the trap that can host the ancillary fermions, yet to be created, close to $\mathbf{y}$, e.g., move the minimum of the laser potential that traps the ancillary atoms there.
 	\item Swap each of the fermionic modes created by $\psi^{\dagger}_m\left(\mathbf{y}\right)$ with that of $\chi^{\dagger}_m$, that is, act with the swap unitaries:
 	\begin{equation}
 	\mathcal{U}_{\text{S}}\left|\Psi\right\rangle\equiv\underset{m}{\prod}\mathcal{U}_{\text{S}}\left(\psi_m\left(\mathbf{y}\right),\chi_m\right) \left|\Psi\right\rangle
 	\end{equation}
 	We note that
 	\begin{equation}
	\mathcal{M}^S\equiv\mathcal{U}_{\text{S}}\mathcal{M}\mathcal{U}^{\dagger}_{\text{S}}=
	\psi^{\dagger}_m\left(\mathbf{x}\right) \left(\underset{\ell =1,...,L}{\prod}U\left(\ell\right)\right)_{mn}\chi_n
 	\end{equation}
 	Due to the swap, in the transformed state $\mathcal{U}_{\text{S}}\left|\Psi\right\rangle$ there are no physical fermions at the vertex $\mathbf{y}$:
 	$\psi_m\left(\mathbf{y}\right)\mathcal{U}_{\text{S}}\left|\Psi\right\rangle=0$ for every component $m$.
 	\item Move the ancilla to the closest link on the string $\mathcal{L}$, and interact with its gauge field Hilbert space using $\mathcal{U}_{G}\left(\ell=L\right)$. This shortens the string by one link:
 	\begin{equation}
 	\begin{aligned}
 	&\mathcal{U}_{G}\left(\ell=L\right)\mathcal{M}^S\mathcal{U}^{\dagger}_{G}\left(\ell=L\right)
 	\\&=\psi^{\dagger}_m\left(\mathbf{x}\right) \left(\underset{\ell =1,...,L-1}{\prod}U\left(\ell\right)\right)_{mn}\chi_n
 	\end{aligned}
 	\end{equation}
 	\item Move the ancilla one link further along $\mathcal{L}$, and let it interact with it, using $\mathcal{U}_{\text{G}}$. Repeat it for all the links until $\ell=1$, where the string is completely removed from $\mathcal{M}$: for $\hat{\mathcal{U}}_{\text{G}}=\mathcal{U}_{G}\left(\ell=1\right)\cdots\mathcal{U}_{G}\left(\ell=L\right)$, we obtain that $\mathcal{M}$ transforms to a sum of spin raising operators:
	\begin{equation}
	\hat{\mathcal{U}}_{G}\mathcal{M}^S\hat{\mathcal{U}}^{\dagger}_{G}
	=\psi^{\dagger}_m\left(\mathbf{x}\right) \chi_m = \underset{m}{\sum}S_{+}\left(m\right)
	\end{equation}
 	where 
	$S_{+}\left(m\right) = \psi^{\dagger}_m\left(\mathbf{x}\right)\chi_m$.
\end{enumerate}
	
	  \begin{figure}
		\includegraphics[width=\columnwidth]{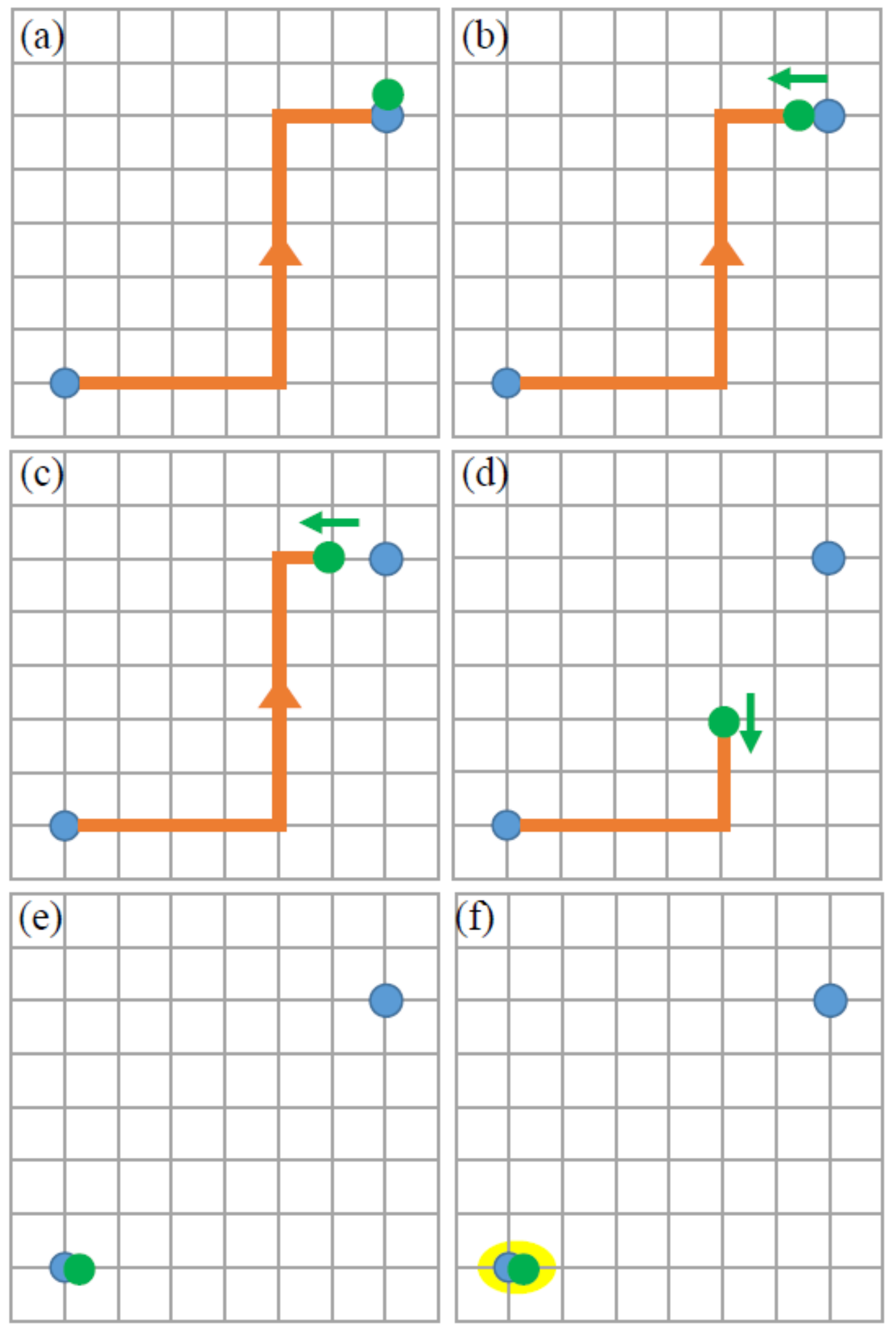}
		\caption{The scheme for a mesonic string: (a) the ancilla (green circle), in its vacuum state, is swapped with the endpoint fermions. 
			(b) the ancilla interacts with the last link, through the de-gauging transformation $\mathcal{U}_{\text{G}}\left(\ell=L\right)$, which will shorten the string by removing this link from it. The ancilla is moved along the string, against its direction, and interacts similarly which each and every link along the way, shortening the string on the way to the beginning, as shown in (c) and (d). In (e) the ancilla is at the starting point, with no string at all. There, one can act locally in a way that is equivalent to acting with a string, or, instead, combine the beginning fermions with the ancilla to a spin as in (f) and rotate it for measurement.}
		\label{Fig3}
	\end{figure}
	
  	For measuring, move the ancilla close to the string's beginning $\mathbf{x}$, and let it interact with its fermionic modes with rotations $\mathcal{U}_{\text{R}}$ for each component - that is, $\hat{U}_{\text{R}}=\underset{m}{\prod}\hat{U}_{\text{R}}^{\left(m\right)}$.
If we denote the complete unitary sequence by $\mathcal{U}_{\text{M}}=\hat{U}_{\text{R}}\hat{\mathcal{U}}_{G}\mathcal{U}_{\text{S}}$, we obtain that
\begin{equation}
\mathcal{U}_{\text{M}} M \mathcal{U}^{\dagger}_{\text{M}} = \underset{m}{\sum}\left(n_{\psi}\left(m\right)-n_{\chi}\left(m\right)\right)
\end{equation}
Implying that
\begin{equation}
\left\langle \psi \right| M \left| \psi \right\rangle =
\underset{m}{\sum}\left\langle \Psi \right| \mathcal{U}^{\dagger}_{\text{M}} \left(n_{\psi}\left(m\right)-n_{\chi}\left(m\right)\right) \mathcal{U}_{\text{M}} \left| \Psi \right\rangle
\end{equation}
- the expectation value of $M$ may be obtained from simple (and local) fermionic number measurements, of a transformed state that is obtained by acting on $\left|\Psi\right\rangle = \left|\psi\right\rangle \otimes \left|\Omega_{\chi}\right\rangle$ by a sequence of local two-body interactions.

The whole process is shown in Fig. \ref{Fig3}.
 
 As in the Wilson loop case, here too we can skip the irreversible  step of measurement (and the rotation) and use the procedure for exciting a meson: we only need the unitary $\mathcal{U}'_{\text{M}}=\hat{\mathcal{U}}_{G}\mathcal{U}_{\text{S}}$, to be used in
 \begin{equation}
 \left(M\left(\mathbf{x},\mathbf{y};\mathcal{L}\right)\left|\psi\right\rangle\right) \otimes \left|\Omega_{\chi}\right\rangle =
  \mathcal{U}_{M}^{'\dagger} \left(\psi^{\dagger}_m\left(\mathbf{x}\right) \chi_m + \chi^{\dagger}_m\psi_m\left(\mathbf{x}\right)\right) \mathcal{U}'_{M} \left|\Psi\right\rangle
 \end{equation}

As in the Wilson loop case, multiple strings can be studied in parallel, using several ancillas. However, unlike the Wilson loop case, since mesonic strings that share endpoints do not commute, once a vertex is used for one meson no other meson emanating from it can be studied in parallel. 
 
\section{Demonstration for $\mathbb{Z}_2$}
 
As an example, we discuss the simplest case (in terms of Hilbert spaces), of a $\mathbb{Z}_2$ lattice gauge theory. It is an Abelian group, and thus each vertex may be occupied by one matter fermion at most, created by the fermionic operator $\psi^{\dagger}\left(\mathbf{x}\right)$. On each link, the gauge field Hilbert space has dimension $2$, and these two-level systems may be thought of simply as qubits. The group element operator is simply
\begin{equation}
U = U^{\dagger}=\sigma_x
\end{equation}

\subsubsection{Wilson Loops}
A Wilson loop operator is simply a product of $\sigma_x$ operators belonging to the links along the path. For Abelian groups, ordering is not important, and no trace is required. Furthermore, in the $\mathbb{Z}_2$ case, since $U=\sigma_x$ is Hermitian, the orientation of a link along the path $\mathcal{C}$ has no meaning. Therefore,
\begin{equation}
\mathcal{W}\left(\mathcal{C}\right) = \underset{\ell \in \mathcal{C}}{\prod}\sigma_x\left(\ell\right)
\end{equation}
The ancilla will simply be another qubit, that we prepare in the initial state
\begin{equation}
\left|\tilde{e}\right\rangle = \left|\tilde{\uparrow}_x\right\rangle = \frac{1}{\sqrt{2}}\left(\left|\tilde{\uparrow}\right\rangle+\left|\tilde{\downarrow}\right\rangle\right)
\end{equation}
(since $\tilde{U}\left|\tilde{e}\right\rangle=\left|\tilde{e}\right\rangle$).

The entangling operation which each link simply takes the form
\begin{equation}
\mathcal{U}_{\text{W}}\left(\ell\right) =\mathcal{U}^{\dagger}_{\text{W}}\left(\ell\right)= \left|\uparrow_x\right\rangle\left\langle\uparrow_x\right|_{\ell} \otimes \tilde{\mathbf{1}} +
\left|\downarrow_x\right\rangle\left\langle\downarrow_x\right|_{\ell} \otimes \tilde{\sigma}_z
\end{equation}

The Wilson loop is measured through
\begin{equation}
\left\langle\psi\right| \mathcal{W}\left(\mathcal{C}\right) \left|\psi\right\rangle =
\left\langle\Psi\right| \mathcal{U}_{\text{W}} \tilde{\sigma}_x \mathcal{U}_{\text{W}}\left|\Psi\right\rangle
\end{equation}
 
 \subsubsection{Mesons}
 
 A mesonic operator will take the form
\begin{equation}
\mathcal{M}\left(\mathbf{x},\mathbf{y};\mathcal{L}\right) = \psi^{\dagger}\left(\mathbf{x}\right) \underset{\ell \in \mathcal{L}}{\prod}\sigma_x\left(\ell\right)\psi\left(\mathbf{y}\right)
\end{equation}
 
 Since there is only one fermionic species, the ancilla will contain only one as well, therefore, both the swap and the rotation operators will not involve any product over species. The local interactions of the ancilla with the gauge field will take the form
 \begin{equation}
 \mathcal{U}_{G}\left(\ell\right) = \exp\left(i \pi n_{\chi} \left(1-\sigma_x\left(\ell\right)\right)/2\right)
 \end{equation}
 and using $\mathcal{U}_{\text{M}} = \mathcal{U}_{\text{R}}\underset{\ell \in \mathcal{L}}{\prod}\mathcal{U}_{\text{G}}\left(\ell\right) \mathcal{U}_{\text{S}}$, we obtain
\begin{equation}
\left\langle \psi \right| M \left| \psi \right\rangle =
\left\langle \Psi \right| \mathcal{U}^{\dagger}_{\text{M}} \left(n_{\psi}-n_{\chi}\right) \mathcal{U}_{\text{M}} \left| \Psi \right\rangle
\end{equation}
 
\section{Summary}
In this work, we demonstrated how to use the stator formalism \cite{reznik_remote_2002,zohar_half_2017} and its application to lattice gauge theories \cite{zohar_digital_2017,zohar_digital_2017-1,bender_digital_2018} to develop a measurement scheme for nonlocal many-body gauge invariant operators - the Wilson loop and the mesonic string. The schemes consist of two-body local interactions with a moving ancilla, that absorbs all the relevant information and then can be read out locally. The scheme may also be used for state preparation, if the last step of the actual measurement is not carried out. 

The method introduced and described here can hopefully be used as a conventional way to extract physical information from the states studied in quantum simulators of lattice gauge theories, which are nowadays becoming a reality as a new non-perturbative tool for studying quantum chromodynamics.

\section*{Acknowledgements}
EZ would like to thank Julian Bender, David B. Kaplan, Martin J. Savage and Jesse R. Stryker for inspiring discussions.

\bibliography{ref}
\end{document}